\documentclass[12pt]{article}
\usepackage{amstex,amssymb}

\newfont{\zapfxii}{eusm10 scaled\magstep1}

\newcommand{\al}{\alpha}
\newcommand{\be}{\beta}
\newcommand{\ga}{\gamma}
\newcommand{\de}{\delta}
\newcommand{\ep}{\epsilon}
\newcommand{\la}{\lambda}
\newcommand{\om}{\omega}
\newcommand{\ph}{\varphi}
\newcommand{\si}{\sigma}

\newcommand{\De}{\Delta}
\newcommand{\La}{\Lambda}
\newcommand{\Si}{\Sigma}

\newcommand{\D}{{\cal D}^1}
\newcommand{\M}{{\cal M}}
\newcommand{\N}{{\cal N}}
\newcommand{\cS}{{\cal S}}
\newcommand{\V}{{\cal V}}

\def\fa{{\mathfrak a}}
\def\fg{\mathfrak g}
\def\fgh{\hat\fg}
\def\fh{\mathfrak h}
\newcommand{\fd}{{\mathfrak d}^1}
\newcommand{\fde}{{\mathfrak d}^1_0}
\newcommand{\fl}{\mathfrak l}
\newcommand{\fs}{\mathfrak s}
\newcommand{\fse}{{\mathfrak s}_0}
\newcommand{\fso}{{\mathfrak s}_1}
\newcommand{\fm}{{\mathfrak m}}
\newcommand{\fn}{{\mathfrak n}}
\newcommand{\fv}{{\mathfrak v}}
\newcommand{\fD}{{\mathfrak D}}
\newcommand{\fL}{{\mathfrak L}}
\newcommand{\fLe}{\fL_0}
\newcommand{\fLo}{\fL_1}
\newcommand{\fN}{\mathfrak N}
\def\fS{\mathfrak S}

\def\fsl{\mathfrak{sl}}
\def\osp#1{\mathfrak{osp}(#1)}

\newcommand{\CC}{{\mathbb C}}
\newcommand{\ZZ}{{\mathbb Z}}

\def\fLb{\,\overline{\!\fL}{}}
\def\fNb{\,\overline{\!\fN}{}}
\def\Tb{\,\overline{\!T}{}}
\def\zb{\bar z}
\def\Tt{\tilde T}
\def\Ut{\tilde U}
\def\jt{\tilde\jmath}
\def\kt{\tilde k}
\def\mut{\tilde\mu}

\newcommand{\Dt}{\partial _\theta}
\newcommand{\Dz}{\partial _z}

\newcommand{\lan}{\langle}
\newcommand{\ran}{\rangle}
\newcommand{\iso}{\simeq}
\newcommand{\ra}{\rightarrow}
\newcommand{\Ra}{\Rightarrow}

\newcommand{\C}{\operatorname{C^\infty}}
\newcommand{\diag}{\operatorname{diag}}
\newcommand{\End}{\operatorname{End}}

\newcommand{\sch}{Schr\"o\-ding\-er}
\newcommand{\qes}{quasi-exactly solvable}
\newcommand{\fdim}{fi\-nite-di\-men\-sion\-al}
\newcommand{\dfo}{differential operator}
\newcommand{\QED}{\quad Q.E.D.}

\newtheorem{thm}{\bf Theorem}[section]
\newtheorem{lemma}[thm]{\bf Lemma}
\newtheorem{prop}[thm]{\bf Proposition}
\newtheorem{cor}[thm]{\bf Corollary}

\numberwithin{equation}{section}

\newcommand{\bpm}{\begin{pmatrix}}
\newcommand{\epm}{\end{pmatrix}}
\def\ni{\noindent}
\def\vr{\vrule width0pt height12pt depth9pt}

\newcounter{mylc}
\renewcommand{\themylc}{\roman{mylc}}
\newenvironment{mylist}{\begin{list}{\themylc )}
{\usecounter{mylc}\settowidth{\labelwidth}{iiii)}}}{\end{list}}

\title{\Large\bf{Quasi-Exactly Solvable Lie Superalgebras\\
       of Differential Operators}%
       \thanks{Supported in part by DGICYT Grant PB95--0401.}}
\author{Federico Finkel\\Artemio Gonz\'alez-L\'opez\\ Miguel A. Rodr\'\i guez\\
\\\em Departamento de F\'\i sica Te\'orica II\\
\em Universidad Complutense de Madrid\\
\em 28040 Madrid, SPAIN}

\date{February 14, 1997}

\begin{document}
\maketitle
\begin{abstract}
In this paper, we study Lie superalgebras of $2\times 2$ matrix-valued
first-order \dfo s on the complex line. We first completely classify all such
superalgebras of finite dimension. Among the \fdim{} superalgebras whose odd
subspace is nontrivial, we find those admitting a
\fdim{} invariant module of smooth vector-valued functions, and classify all the
resulting
\fdim{} modules. The latter Lie superalgebras and their modules are the
building blocks in the construction of QES quantum mechanical models for spin
$1/2$ particles in one dimension.
\end{abstract}
\newpage

\section{Introduction}\label{sec.intro}

The discovery of \qes{} (QES) spectral problems over the past decade has
been a continuous source of interesting mathematical problems. The characteristic
feature of a QES Hamiltonian is that a nontrivial portion of its spectrum, but
not necessarily all of it, can be computed algebraically. Thus, QES spectral problems
occupy an intermediate position between exactly
solvable problems, whose spectrum can be completely described, and the vast
majority of non-solvable ones. Lie algebras of \dfo s have been used extensively to
generate physically interesting QES \sch{} operators, \cite{ST89},
\cite{Tu88}, \cite{Us89}; see also the review book~\cite{Us94}. The basic idea underlying the
application of Lie algebras of \dfo s to constructing QES models can be summarized as follows:
if $\fg$ is a Lie algebra of first-order \dfo s with an invariant \fdim{}
$\fg$-module of smooth fuctions $\N$, then any scalar Hamiltonian
$H=-\De+V$ which can be expressed as a quadratic combination in the generators
of $\fg$,
\begin{equation}\label{quadratic}
H=\sum_{a,b}\,c_{ab}T^aT^b+\sum_a\,c_aT^a+c_0\,,\qquad T^a\in\fg\,,
\end{equation}
will automatically preserve $\N$. Consequently, if the functions in $\N$
satisfy suitable boundary conditions,
one can compute $\dim\N$ eigenfunctions and eigenvalues by diagonalizing
the finite matrix which represents $H$ in $\End\N$. Therefore, the classification
under some well-adapted notion of equivalence
of all \fdim{} Lie algebras of first-order \dfo s admitting an
invariant module of functions (henceforth called {\it QES Lie algebras\/})
is a good starting point to obtain large families of QES Hamiltonians.
In the one-dimensional scalar case the classification is very simple.
Indeed, every \fdim{} QES Lie algebra in one real or complex variable is isomorphic
to a subalgebra of (a central extension of) $\fsl_2$, \cite{Mi68}, \cite{Tu88},
\cite{KO90}. The classification of \fdim{} QES Lie algebras in two variables is
considerably more involved. There are several inequivalent families of QES Lie
algebras, some of them of arbitrary dimension, and the real and complex
classifications no longer coincide, \cite{GKO91}, \cite{GKO96}.

The above classification is not, however, the end of the problem. One still
has to determine the conditions under which a quadratic combination of the
form~\eqref{quadratic} is equivalent to a \sch{} operator $-\De+V$. In the one-dimensional
case, it turns out that any quadratic combination~\eqref{quadratic} may be written (locally)
in \sch{} form by the combination of a change of the independent variable and a
gauge transformation with a non-vanishing function. The situation in higher dimensions
is again more complicated. Explicit necessary and sufficient conditions for the
equivalence under local diffeomorphisms and gauge transformations of scalar
second-order \dfo s were first found by \'E.~Cotton, \cite{Co00}. As a special
case, one obtains conditions for the equivalence of a second-order \dfo{} to a
\sch{} operator acting on a (in general) curved space-time. These conditions
have been solved only in some particular cases, and appear to be too complicated
to be solved in full generality; see \cite{Mi95} for an in-depth study.

This formalism may be extended to deal with matrix-valued differential equations,
suitable for the description of the dynamics of particles with nonzero spin,
\cite{ST89}, \cite{BK94}, \cite{FGR96}, or the treatment of coupled-channel scattering
problems, \cite{Zh96}. The procedure for constructing matrix-valued QES Hamiltonians
is essentially the same as in the scalar case, with the role of the Lie algebra $\fg$ 
now being played by a {\em Lie superalgebra} $\fS$ of matrix-valued \dfo s
with an invariant subspace of vector-valued functions, \cite{FGR96}. In principle, $\fS$ need not
be \fdim{}; in practice, however, the only examples constructed so far with this
method are associated to \fdim{} Lie superalgebras, \cite{ST89}, \cite{BK94}, \cite{FGR96}.
Lie superalgebras
of \dfo s are
significantly less understood than ordinary Lie algebras. In fact, to the best of the
authors' knowledge, no general classification of \fdim{} Lie superalgebras
of \dfo s has ever been attempted. The goal of our paper is precisely that of
classifying all QES Lie superalgebras of $2\times 2$ matrix first-order \dfo s
in one complex variable. The Lie superalgebras thus obtained can be readily used to
construct new examples of second-order $2\times 2$ matrix-valued QES operators, by
taking quadratic combinations in the generators of $\fS$ and
performing a suitable change of the independent variable and/or a gauge transformation.
Necessary and sufficient conditions for the equivalence of a $2\times 2$
matrix-valued \dfo{} to a \sch{} operator were obtained in \cite{FGR96} and
\cite{FK96}. A number of QES Lie superalgebras preserving a two-component vector-valued module
of polynomials in two complex variables were recently studied in \cite{FK96b}.

The paper is organized as follows. In Section~2 we summarize
the main results concerning Lie algebras of first-order scalar \dfo s on the
complex line. In Section~3 we
outline our classification scheme for the \fdim{}
Lie superalgebras $\fS$ of $2\times 2$ matrix-valued first-order \dfo s. We start
with a basic result describing the structure of the even and odd subspaces of $\fS$.
The classification of all such Lie superalgebras $\fS$ is then shown
to be completely equivalent to classifying the \fdim{} graded subalgebras
$\fs$ of the Lie superalgebra $\fd$ of first-order \dfo s in one ordinary
variable and one Grassmann variable taking values in the one-generator
Grassmann algebra $\La^1$. In Section~4 we classify all possible even
subalgebras $\fl$ of $\fd$. The concept of a translation bimodule
introduced in this section turns out to play an essential role
in the classification, as later shown in Sections~5 and 6. Section~5 is devoted to
the classification of the odd subspaces $\fso$ corresponding to each subalgebra $\fl$
obtained in Section~4. We first state a necessary condition for $\fl$
to admit a nonzero odd subspace $\fso$.
The list of the even subalgebras $\fse$ and their corresponding
nontrivial odd subspaces $\fso$ is then presented in Tables~1--4. Finally,
in Section~6 the classification is completed by finding all the Lie superalgebras with
nontrivial odd subspace that admit a \fdim{} module
$\fn\subset\C(\La^1)$. The associated \fdim{} modules are also classified, and the
results are summarized in Tables~5--9.

\section{Lie algebras of \dfo s.}

In this section, we briefly review the basic theory of Lie algebras of
first-order scalar differental operators on the complex line, which will 
serve as a helpful guide in what follows.

Let $\D$ denote the Lie algebra of \dfo s of the form
\begin{equation}\label{1dfo}
T=f(z)\Dz+g(z)\,,
\end{equation}
where $f$ and $g$ are analytic functions of a complex variable $z$,
and the Lie bracket is given by the usual commutator. There are two
pseudogroups of transformations acting naturally on $\D$ which preserve its
Lie algebra structure, namely local diffeomorphisms $\zb=\ph(z)$
and gauge transformations by a non-vanishing function $u(z)$. The action of these
transformations on an operator $T$ is given by
$$
T\mapsto\Tb,\qquad\hbox{\rm with}\qquad
\Tb(\zb)=u(z)\cdot T(z)\cdot u(z)^{-1}\,.
$$
We shall say that two Lie subalgebras of $\D$ are {\em equivalent} if they
can be mapped into each other by a {\em fixed} combination of a local
diffeomorphism and a gauge transformation.

The Lie algebra $\D$ splits
naturally into the semi-direct product of
the subalgebras of vector fields $\V$ and multiplication operators $\M$:
$$
\D=\V\ltimes\M\,.
$$
The classification up to a change of variable of the \fdim{} subalgebras of $\V$
is due to S. Lie, \cite{Li80}:
\begin{lemma}\label{lem.Lie}
Every nonzero \fdim{} Lie algebra of vector fields on the line is related
by a local change of variable to one of the following Lie algebras:
$$
\fh^1=\lan\Dz\ran\,,\qquad\fh^2=\lan\Dz,z\Dz\ran\,,
\qquad\fh^3=\lan\Dz,z\Dz,z^2\Dz\ran\,.
$$
\end{lemma}
Since the natural projection $\pi:\D\ra\V$ mapping a \dfo{}
\eqref{1dfo} to its vector field part $f(z)\Dz$ defines a Lie
algebra homomorphism, and gauge transformations leave the vector field
part unaffected, one may use Lie's classification of vector fields
to derive the classification of all \fdim{} subalgebras
of $\D$, \cite{Mi68}, \cite{KO90}:
\begin{thm}\label{thm.Miller}
Let $\fg$ be a \fdim{} subalgebra of $\D$. Then $\fg$ is equivalent to one of
the following Lie algebras:
\begin{mylist}
\item $\fg^0=\lan g_i(z)\:|\:1\leq i\leq m\ran$, where the functions $g_i$ are linearly
independent.
\item $\fg^1=\lan\Dz,z^ie^{\mu z}\:|\:0\leq i\leq m_\mu,\;\mu\in M\ran$. Here
$M$ denotes a finite collection of complex numbers.
\item $\fg^2=\lan\Dz,z\Dz,z^i\:|\:0\leq i\leq m\ran$.\\[.2cm]
      $\fgh^2=\lan\Dz,z\Dz+\al\ran$, where $\al\in\CC$.
\item $\fg^3=\lan\Dz,z\Dz,z^2\Dz+2\al z,1\ran$, where $\al\in\CC$.\\[.2cm]
      $\fgh^3=\lan\Dz,z\Dz+\al,z^2\Dz+2\al z\ran$, where $\al\in\CC$.
\end{mylist}
\end{thm}

\section{Lie superalgebras of \dfo s}

We now focus our attention on $2\times 2$ matrix-valued Lie
superalgebras of first-order \dfo s.
Let $\fD$ denote the associative algebra of all
$2\times 2$ matrix \dfo s on a complex variable $z$. We introduce
a $\ZZ_2$-grading in $\fD$ in the usual way: an operator
$$
T=\bpm a & b \\ c & d \epm\,,
$$
where $a$, $b$, $c$, and $d$ are scalar \dfo s, is said to be {\em even} if
$b=c=0$, and {\em odd} if $a=d=0$. This grading, combined with the
generalized Lie product
$$
[A,B]_s=AB-(-1)^{\deg A\deg B}BA\,,
$$
endows $\fD$ with a Lie superalgebra structure. We shall be interested in \fdim{} graded
subalgebras of the graded subspace $\fD^{(1)}\subset\fD$ of first-order
\dfo s. Two such graded subalgebras $\fL$ and $\fLb$ will be considered {\em equivalent}
if their elements $T\in\fL$ and $\Tb\in\fLb$  are related by a {\em fixed}
local change of variable $\zb=\ph(z)$ and a gauge transformation consistent
with the grading:
\begin{equation}\label{equiv}
T\mapsto\Tb,\qquad\hbox{\rm with}\qquad
\Tb(\zb)=U(z)\cdot T(z)\cdot U(z)^{-1}\,,
\end{equation}
where the $2\times2$ invertible complex matrix $U(z)$ is either diagonal or antidiagonal.
This is a very natural notion of equivalence in the context of QES problems; indeed, if
$\fN$ is an $\fL$-module of vector-valued functions and
$\fLb$ is equivalent to $\fL$ under the mapping~\eqref{equiv}, then
$\fNb=U\cdot\fN$ is an invariant module for $\fLb$.
The first aim of this paper consists in classifying under the above equivalence
all \fdim{} graded subalgebras of $\fD$ contained in $\fD^{(1)}$ .

We begin with the following elementary result, that we shall state without proof:
\begin{lemma}\label{lem.DD1}
Let $\fL\subset\fD^{(1)}$ be a graded subalgebra of $\fD$, and let $\fLe$ and 
$\fLo$ denote its even and odd subspaces, respectively. We then have:
\begin{mylist}
\item 
Either all the elements of $\fLo$ are of the form
\begin{equation}\label{T1}
T_1=\bpm 0 & \phi\Dz+\om \\ \chi & 0 \epm\,,
\end{equation}
or all its elements are of the form
\begin{equation}\label{TT1}
\Tt_1=\bpm 0 & \chi \\ \phi\Dz+\om & 0 \epm\,,
\end{equation}
where $\phi$, $\om$, and $\chi$ are analytic functions of $z$.
\item If $\fLo$ is nonzero, the elements of $\fLe$ are of the form:
\begin{equation}\label{T0}
T_0=\bpm f\Dz+h_1 & 0 \\ 0 & f\Dz+h_2 \epm\,,
\end{equation}
where $f$, $h_1$, and $h_2$ are analytic functions of $z$.
\end{mylist}
\end{lemma}
Let us denote by $\fD^1$ (respectively $\tilde\fD^1$) the graded subalgebra
of $\fD$ generated by all \dfo s of the form $T_0$ and $T_1$ (respectively $T_0$ and
$\Tt_1$) in \eqref{T0} and \eqref{T1} (respectively \eqref{T0} and \eqref{TT1}). The
graded subalgebras $\fD^1$ and $\tilde\fD^1$ are equivalent, since they are related
by a gauge transformation with constant matrix
$$
\Ut=\bpm 0 & 1 \\ 1 & 0 \epm\,.
$$
Furthermore, the gauge transformations preserving $\fD^1$ (or
$\tilde\fD^1$) are generated by diagonal matrices $U(z)=\diag(\al,\be)$, where $\al$ and $\be$ are
non-vanishing analytic functions of $z$. Therefore, without any loss of generality,
we can limit ourselves to classifying the \fdim{} graded
subalgebras of $\fD^1$ under local diffeomorphisms and gauge transformations
by diagonal matrices.

There is an alternative formalism for describing Lie superalgebras of
\dfo s which makes use of Grassmann variables. The matrix classification
scheme described above is completely equivalent to classifying
all \fdim{} subalgebras $\fs$ of the Lie superalgebra $\fd$ of first-order
\dfo s in one ordinary variable $z$ and one Grassmann variable $\theta$
which take values in the one-generator Grassmann algebra $\La^1$. The even
and odd subspaces of $\fd$ are generated by \dfo s of the form
\begin{align}
T_0 &= f(z)\Dz+g(z)\theta\Dt+h(z)\,,\label{even}\\
\intertext{and}
T_1 &= \theta\phi(z)\Dz+\chi(z)\Dt+\om(z)\theta\,,\label{odd}
\end{align}
respectively, where $f$, $g$, $h$, $\phi$, $\chi$, and $\om$ are analytic functions of $z$.
The appropriate equivalence transformations are in this case
changes of the independent variables preserving the relation
$\{\Dt,\theta\}=1$, namely
\begin{equation}\label{civ}
\zb=\ph(z)\,,\qquad \bar\theta=\be(z)\theta\,,
\end{equation}
and gauge transformations with a gauge factor of the form $u=\al(z)$, where
$\al$ and $\be$ are non-vanishing analytic functions of $z$. The standard
identifications
$$
\theta\leftrightarrow\si^+\,,\qquad \Dt\leftrightarrow\si^-\,,\qquad
\text{with}\quad\si^+=(\si^-)^t=\bpm 0 & 1 \\ 0 & 0\epm\,,
$$
lead directly to the equivalence of both formalisms. For the sake
of simplicity, we shall use in what follows the Grassmann variable notation.
Our first step will thus be to classify all \fdim{} Lie subalgebras
$\fl$ of $\fde$, the even subspace of $\fd$. We will then restrict ourselves to the
subalgebras $\fse\subset\fde$ which admit a nontrivial \fdim{} odd subspace $\fso{}$ satisfying
the conditions
\begin{equation}\label{conditions}
[\fse,\fso]\subset\fso\,,\qquad \{\fso,\fso\}\subset \fse\,.
\end{equation}
We will determine all such odd subspaces $\fso$, and then
we will obtain the \fdim{} modules of functions
$\fn\subset\C(\La^1)\iso\C(\CC)\otimes\lan 1,\theta\ran$ associated to each
Lie superalgebra $\fs=\fse\oplus\fso$. Note that the $\La^1$-valued function
$f(z)+g(z)\theta$ is identified with the two-component function $(g(z),f(z))^t$
in the matrix formalism.

\section{Lie algebras of even \dfo s}

In this section we classify the \fdim{} subalgebras $\fl$ of the even
subspace $\fde$ of $\fd$.

The Lie algebra $\fde$ admits the following natural decomposition:
$$
\fde=\V\ltimes\fa\,,
$$
where $\fa$ is the abelian Lie algebra of all operators
of the form
$$
\hat T_0=g(z)\theta\Dt+h(z)\,.
$$
We first observe that the projection $\pi:\fde\ra\V$ mapping an even differential
operator of the form~\eqref{even} to $f\Dz\in\V$ is a homomorphism of Lie algebras.
Moreover, both $\fl$ and its transformed under a change of the odd variable,
\begin{equation}\label{civ1}
\zb=z\,,\qquad \bar\theta=\be(z)\theta\,,
\end{equation}
and/or a gauge transformation clearly have the same projection in $\V$.
Therefore, $\pi(\fl)$ is either zero or equivalent under a change of the even
variable
$$
\zb=\ph(z)\,,\qquad \bar\theta=\theta\,,
$$
to one of the three Lie algebras $\fh^i$ in Lemma~\ref{lem.Lie}. The situation
in this respect is completely analogous to the one in the scalar case.
Moreover, the Lie algebra $\fv$ of all vector fields of the form
\begin{equation}\label{vf}
\tilde T_0=f(z)\Dz+g(z)\theta\Dt\,,
\end{equation}
is isomorphic to $\D$, under the identification
$$
\tilde T_0\mapsto f(z)\Dz+g(z)\,.
$$
Under this identification, a gauge transformation in $\D$ by a non-vanishing
function $u=\al(z)$ becomes a change of the odd variable in $\fv$ of the form
$\bar\theta=\theta/\al$.
Thus, the classification of the \fdim{} subalgebras of $\D$ in Theorem~\ref{thm.Miller}
and the classification of the \fdim{} subalgebras of $\fv$ under changes of variables~
\eqref{civ} are identical.

We need to introduce some additional notation at this stage. Let $V=\lan v_1,v_2\ran$ be an
abstract two-dimensional complex vector space. We define a
{\em $V$-trans\-la\-tion bimodule} as a \fdim{} $\fh^1$-invariant subspace
of $\M\otimes V$. The following result provides a detailed description of
$V$-translation bimodules: 

\begin{prop}\label{prop.bimodule}
The most general $V$-translation bimodule is a direct sum
$$
\fm(V,M)=\bigoplus \begin{Sb} \mu\in M \\ i=1,2,3 \end{Sb}
                  \fm^i_\mu(V)\,e^{\mu z}\,,
$$
where $M$ is a finite collection of complex numbers, and
\begin{gather}\label{m123}
\fm^1_\mu(V)=\big\lan z^k\,v_1\:|\:0\leq k\leq m_\mu\big\ran\,,\qquad
\fm^2_\mu(V)=\big\lan z^k\,v_2\:|\:0\leq k\leq n_\mu\big\ran\,,\notag\\
\fm^3_\mu(V)=\big\lan z^{m_\mu+k}\,v_1+z^{n_\mu}\sum_{j=1}^k c^{\mu,k}_j\,z^j\,v_2\,
                  \:|\:1\leq k\leq r_\mu\big\ran\,,
\end{gather}
with
\begin{equation}\label{c}
c^{\mu,k}_j=\frac{(m_\mu+k)!(n_\mu+1)!}{(n_\mu+j)!(m_\mu+k-j+1)!}\,c^{\mu,k-j+1}\,,
\end{equation}
where $c^{\mu,1},\ldots,c^{\mu,r_\mu}\in\CC$, and $c^{\mu,1}\neq 0$. By convention, the 
indices $m_\mu$, $n_\mu$, and $r_\mu$ take the values $-1$, $-1$, and $0$,
respectively, when their corresponding modules $\fm^i_\mu(V)$ are zero.
\end{prop}

\ni{\em Remark.} The notation $\fm^i_\mu(V)$ is an abbreviated notation, since the
latter sets actually depend on $\mu$ through the parameters $m_\mu$, $n_\mu$, $r_\mu$
and $c^{\mu,l}$.

\ni{\em Proof.} The only nontrivial point is the structure of the module
$\fm^3_\mu(V)$ of {\em mixed} vectors. A mixed vector $g\,v_1+h\,v_2$ may
always be chosen to be proportional to
$$
z^{m_\mu+k}e^{\mu z}\,v_1+\sum\begin{Sb}\nu\in N\\1\leq j\leq s_\nu\end{Sb}
   c^{\mu,k}_{\nu,j}\,z^{n_\nu+j}e^{\nu z}\,v_2\,,
$$
with $k=1,\ldots,r_\mu$ and complex numbers $c^{\mu,k}_{\nu,j}$. Acting with
$\Dz$ on each of these vectors for $k=1,\ldots,r_\mu$, we immediately
obtain $c^{\mu,k}_{\nu,j}=\de_{\mu\nu}\,c^{\mu,k}_j$, with $c^{\mu,k}_j$
given by~\eqref{c} for $j\leq k$ and $c^{\mu,k}_j=0$ for $j>k$.\QED\\

From now on, we shall use the following conventions.
The highest value $i_{\max}$ of an index $i$ labeling a collection of
operators $\cS=\{T_i\:|\:i_{\min}\leq i\leq i_{\max}\}$ will be set to
$i_{\min}-1$ whenever $\cS=\emptyset$. The letters $\al,\be,\ga,\de$ will denote
complex numbers, while the letter $c$ will be reserved for a {\em nonzero} complex number.
We shall also define the set $\fm_0(V)$ by
$$
\fm_0(V)=\fm(V,\{0\})\,,\quad\text{with}\quad
c^{0,k}=\de_{k1}\,c\,.
$$
Explicitly,
\begin{equation}\label{m0}
\fm_0(V)=\big\lan z^i\,v_1,\,z^j\,v_2,\,z^k(z^m\,v_1+c_k z^n\,v_2)\big\ran\,,
\end{equation} 
with $0\leq i\leq m\,$, $0\leq j\leq n\,$, $1\leq k\leq r\,$, and
\begin{equation}\label{c0}
c_k=\frac{(m+k)!(n+1)!}{(n+k)!(m+1)!}\,c\,.
\end{equation}
Note that $c_k=c$ if $m=n$.

We are now ready to state the classification theorem for the \fdim{} subalgebras
of $\fde$. In what follows, a semicolon ``;'' will be used to separate those
generators which characterize a given set from those which may or may not be
present, and
$$
\fa_0 = \lan\theta\Dt,1\ran.
$$

\begin{thm}\label{thm.even}
Let $\fl$ be a \fdim{} subalgebra of $\fde$. Then $\fl$ is equivalent to one
of the following Lie algebras:
\begin{mylist}
\item $\fl^0=\lan g_i(z)\theta\Dt+h_i(z)\:|\:1\leq i\leq m\ran$, where the
operators are linearly independent.
\item $\fl^1=\lan\Dz\,;\,\fm(\fa_0,M)\ran$, where
$\fm(\fa_0,M)$ is an $\fa_0$-translation bimodule.
\item $\fl^2=\lan\Dz,\,z\Dz+\al\theta\Dt+\be\,;\,\fm_0(\fa_0)\ran$, where $\fm_0(\fa_0)$
is given by~\eqref{m0}. We also have the constraint $m=n$ if $r\geq 0$.
\item $\fl^3=\lan\Dz,\,z\Dz+\al\theta\Dt+\be,\,z^2\Dz+2\al z\theta\Dt+2\be z\,;\,
\theta\Dt+\ga,\,1\ran$
\end{mylist}
\end{thm}

\ni{\em Remark.} Some of the generators may be simplified depending
on the presence of the optional generators. For instance, we may take
$\al=\be=0$ in $\fl^2$ if $m,n\geq 0$, or the second generator of $\fl^3$
may be taken as $z\Dz+\be$ if $\theta\Dt+\ga$ is present.

\ni{\em Proof.} Case {\em i)} is obvious. Let us start with $\fl^1$. In view
of the isomorphism $\fa\iso\M\otimes\fa_0$, we have:
$$
\fl^1=\lan\Dz+g\theta\Dt+h,\,\fm(\fa_0,M)\ran\,,
$$
where $g$ and $h$ may be eliminated by an appropriate change of the
odd variable~\eqref{civ1} and a gauge transformation. Consider now $\fl^2$.
Commuting $z\Dz+g\theta\Dt+h$ with the elements of $\fm(\fa_0,M)$ we conclude that
$M=\{0\}$. If there are mixed operators present, that is, if $r=r_0\neq 0$
in~\eqref{m123}, then $m=n$ and $c^{0,k}_j=\de_{kj}c$. Furthermore, we may eliminate
$g$ (respectively $h$) from $z\Dz$ by a suitable change of the odd variable
(respectively gauge transformation) unless it is a constant. Finally, we may
also have a generator $z^2\Dz+g\theta\Dt+h$. Commuting it with the generators
in $\fm_0(\fa_0)$ we conclude that $\fl^3\cap\fa$ is a subspace of
$\lan\theta\Dt+\ga,1\ran$. Since
$$
[\Dz,z^2\Dz+g\theta\Dt+h]=2z\Dz+g'\theta\Dt+h'\,,
$$
we conclude that $g=2\al z\theta\Dt$ and $h=2\be z$, for $\al,\be\in\CC$.\QED

\section{The odd subspaces}

We will determine next the possible \fdim{} odd subspaces $\fso$
for each of the families of even Lie algebras $\fl^i$ of Theorem~\ref{thm.even}.
These odd subspaces must verify the commutation relations~\eqref{conditions}. Let us
remark that we are left with very few equivalence transformations preserving
the canonical forms $\fl^i$ listed in Theorem~\ref{thm.even} to simplify the
odd subspaces. In spite of this, we shall see that some of the Lie algebras of
Theorem~\ref{thm.even} admit only trivial \fdim{} odd subspaces.
\begin{lemma}
Let $\fse$ be a subalgebra of $\fd_0$, and assume that $\fso$ is a nontrivial \fdim{}
odd subspace for $\fse$. If $g(z)\theta\Dt+h(z)$ belongs to $\fse$, then $g(z)$ must be a
constant.
\end{lemma}
We thus have:
\begin{cor}\label{cor.even}
Let $\fs$ be a \fdim{} graded subalgebra of $\fd$ with a nontrivial odd
subspace $\fso$. The even subalgebra $\fse$ of $\fs$ is then equivalent to one
of the following Lie algebras:
\begin{mylist}
\item $\fse^0=\lan\ep\theta\Dt+h_1(z)\,;\,h_l(z)\:|\:2\leq l\leq s\ran$, where the
functions $h_l$ are linearly independent.
\item $\fse^1=\lan\Dz\,;\,\ep\theta\Dt+\al z^{s_0+1},\,z^le^{\si z}\:|\:
0\leq l\leq s_\si,\;\si\in\Si\ran$. Here $\Si$ is a finite collection of
complex numbers, $s_0=-1$ if $0\notin\Si$, and $\al=0$ if $\ep=0$.
\item $\fse^2=\lan\Dz,\,z\Dz+\al\theta\Dt+\be\,;\,\ep\theta\Dt+\ga,\,z^l\:|\:
0\leq l\leq s\ran$.
\item $\fse^3=\lan\Dz,\,z\Dz+\al\theta\Dt+\be,\,z^2\Dz+2\al z\theta\Dt+2\be z\,;\,
\ep\theta\Dt+\ga,\,1\ran$.
\end{mylist}
The parameter $\ep$ takes values $0,1$.
\end{cor}
In Tables 1--4, we present the possible odd subspaces corresponding to each family
of Lie algebras $\fse^i$ in Corollary~\ref{cor.even}. It is convenient at this
stage to introduce the following convention: the parameters $\ep$, $\hat\ep$
and $\tilde\ep$ will take the values
$0,1$, and $\ep^\ast=1-\ep$, $\hat\ep^\ast=1-\hat\ep$.

As an illustration of how these tables
were constructed, we shall examine the Lie superalgebras of types $1_3$ and $1_4$ in Table~2.
Let $T_1\in\fso$ be an operator of the
form~\eqref{odd}, with $\phi\neq 0$. If $\chi=0$ for all such operators, then
another operator $\Tt_1=\tilde\chi\Dt+\tilde\om\theta$, with $\tilde\chi\neq 0$,
must be present in $\fso$. Let us assume that $\ep=\al=0$ in $\fse^1$, and
$\chi\neq 0$. Anticommuting $T_1$ with itself we conclude that
$\phi\chi$ is a constant and $\phi\chi'=0$. Thus $T_1=\theta\Dz+c\Dt+\om\theta$
for some constant $c\neq 0$. Since $[\Dz,T_1]=\om'\theta$, we conclude that
$\om'\in\lan z^i e^{\mu z}\:|\:0\leq i\leq m_\mu,\;\mu\in M\ran$.
Therefore, $\om=\de z^{m_0+1}$ for some $\de\in\CC$. Computing
$[z^l e^{\si z},T_1]$ we conclude that 
$m_\si\geq s_\si$ for $0\neq\si\in\Si$ and $m_0\geq s_0-1$. Conversely,
from $\{T_1,z^i e^{\mu z}\theta\}$ we deduce that
$s_\mu\geq m_\mu$ for $0\neq\mu\in M$ and $s_0\geq m_0+1$ (or $s_0\geq m_0$ if $\de=0$).
In any case, we can gauge away the term $\de z^{m_0+1}$ in $T_1$ and rescale $\theta$ so that
$c=1$ without affecting $\fse^1$. We obtain the Lie superalgebra $1_4$ in
Table~2. The choice $\chi=0$ in $T_1$ leads to the Case~$1_3$.

\begin{table}[p]
\caption{Odd subspaces for Lie algebras of type $\fse^0$.}
{\small
\begin{center}
\begin{tabular}{c|l|l} Label & \multicolumn{1}{c|}{$\fso$} &
\multicolumn{1}{c}{Rules} \vr\\
\hline\hline $0_1$ & $\lan\theta(\phi_i\Dz+\om_i),\,h'_l\phi_i\theta\ran$ &
$1\leq i\leq m$\,. \vr\\
\hline $0_2$ & $\lan\chi_i\Dt+\om_i\theta\ran$ &
    $1\leq i\leq m$\,;\qquad $\chi_i\om_i=0$ if $\ep=1$\,;\vr\\
& & $\chi_i\om_k+\chi_k\om_i\in\fse^0$\,.
\end{tabular}
\end{center}}
\end{table}

\begin{table}[p]
\caption{Odd subspaces for Lie algebras of type $\fse^1$.}
\small{
\begin{center}
\begin{tabular}{c|l|l} Label & \multicolumn{1}{c|}{$\fso$} &
\multicolumn{1}{c}{Rules} \vr\\

\hline\hline $1_1$ & $\fm(\lan\theta\Dz,\theta\ran,M)$ &
$n_{\mu+\si}\geq s_\si+m_\mu+r_\mu-\de_{\al 0}\de_{\si 0}\quad$ if \vr\\
& & \qquad {\em i)\:}  $m_\mu\geq 0$ or $r_\mu\geq 1$\,, \\
& & \qquad {\em ii)} $\si\in\Si$, with $\al\neq 0$ if $s_0=0$\,. \vr\\

\hline $1_2$ & $\fm(\lan\Dt,\theta\ran,M)$ & $r_\mu=0$ if $\ep=1$\,.\vr\\
& & $m_\mu,\,n_{\mut}\geq 0\,\Ra\,s_{\mu+\mut}\geq m_\mu+n_{\mut}$\,.\\
& & $r_\mu\geq 1,\:n_{\mut}\geq 0\,\Ra\,s_{\mu+\mut}\geq
    m_\mu+r_\mu+n_{\mut}$\,.\vr\\
& & $r_\mu\geq 1,\:m_{\mut}\geq 0\,\Ra\,s_{\mu+\mut}\geq
    n_\mu+r_\mu+m_{\mut}$\,.\\
& & $r_\mu,r_{\mut}\geq 1,\:
    m_\mu+n_{\mut}\neq m_{\mut}+n_\mu$\vr\\
& & \qquad $\Ra\,s_{\mu+\mut}\geq
    r_\mu+r_{\mut}+\max(m_\mu+n_{\mut},m_{\mut}+n_\mu)$\,.\\
& & $r_\mu,r_{\mut}\geq 1,\:m_\mu+n_{\mut}=m_{\mut}+n_\mu$,\:
    $S^p_{\mu\mut}\neq\emptyset$ for some $p$\vr\\
& & \qquad $\Ra\,s_{\mu+\mut}\geq m_\mu+n_{\mut}+
    \underset{S^p_{\mu\mut}\neq\emptyset}{\max}\,p$\,.\vr\\

\hline $1_3$ & $\lan\theta\Dz,\,\Dt+\hat\ep z^{m_0+1}\theta\,;\,z^ie^{\mu z}\theta\ran$ &
$\hat\ep=1\Ra\ep=\al=0$\,.\vr\\
& & $\Si^\ast=M^\ast,\quad s_\mu=m_\mu$\,. Either\\
& & \qquad {\em i)\;} $\hat\ep=\al=0\Ra s_0=m_0,\,m_0+1$\,.\vr\\
& & \qquad {\em ii)\:} $\hat\ep=0$, $\al\neq 0\Ra s_0=m_0$\,.\\
& & \qquad {\em iii)} $\hat\ep=1\Ra s_0=m_0+1$\,.\vr\\

\hline $1_4$ & $\lan\theta\Dz+\Dt\,;\,z^ie^{\mu z}\theta\ran$ &
$\ep=\al=0$\,.\vr\\
& & $\Si^\ast=M^\ast,\quad s_\mu=m_\mu,\quad s_0=m_0,\,m_0+1$\,.\vr\\
\end{tabular}
\end{center}}\vspace{.5cm}
\footnotesize{The set $S^p_{\mu\mut}\subset\ZZ^4$ in Case $1_2$ is defined as
$$
S^p_{\mu\mut}=\big\{(k,\kt,j,\jt)\:|\:k+\jt=\kt+j=p\;\text{and}\;
c^{\mu,k}_j+c^{\mut,\kt}_{\jt}\neq 0\,;\;
1\leq j\leq k\leq r_{\mu}\,,\; 1\leq\jt\leq\kt\leq r_{\mut}\,\big\}
$$
In Cases $1_3$ and $1_4$ the index $i=0,\ldots,m_\mu$, $\mu\in M$.}
\end{table}

\begin{table}[h]
\caption{Odd subspaces for Lie algebras of type $\fse^2$.}
\small{
\begin{center}
\begin{tabular}{c|l|l} Label & \multicolumn{1}{c|}{$\fso$} &
\multicolumn{1}{c}{Rules}\vr\\

\hline\hline $2_1$ & $\fm_0(\lan\theta\Dz,\theta\ran)$ &
$r\geq 1\,\Ra\,m=n+1$\: and $\:s=-1,0$\,.\vr\\
& & $r=0,\;s>0,\;m\geq 0\,\Ra\,n\geq m+s-1$\,.\vr\\

\hline $2_2$ & $\fm_0(\lan\Dt,\theta\ran)$ &
$r=0$ if $\ep=1$\,.\vr\\
& & $r\geq 1\,\Ra\,m=n+2\al;\;\al$ semi-integer $\geq-(n+1)/2$\,.\\
& & $s\geq m+n+2r$\: if $\:m,n\geq 0$\: or $\:r\geq 1$\,.\vr\\

\hline $2_3$ & $\lan\theta\Dz,\,\Dt+\hat\ep z^{m+1}\theta;\,z^i\theta\ran$ &
$\hat\ep=1\,\Ra\,s=m+1=-2\al$\: and $\:\ep=0$\,.\vr\\
& & $\hat\ep=0\,\Ra\,s=m,m+1$\,.\vr\\

\hline $2_4$ & $\lan\theta(z\Dz+\de),\,\theta\Dz,\,\Dt+\ep^\ast\hat\ep\,\theta;\,z^i\theta\ran$ &
$\al=0$\: and either:\vr\\
& & \qquad {\em i)\:} $s=m$\,,\; $\de=\be$\,,\; and $\hat\ep=0$\,.\\
& & \qquad {\em ii)\:} $s=m+1=0$\,.\vr\\

\hline $2_5$ & $\lan\theta\Dz+\hat\ep^\ast z\Dt,\,\hat\ep z\Dt,\,\Dt;\,z^i\theta\ran$ &
$s=m+1$\: and either:\vr\\
& & \qquad {\em i)\:} $\ep=\be=0$\,,\; $\al=1$\,.\\
& & \qquad {\em ii)\:} $\ep=\hat\ep=1$\,,\; $\be=-\ga$\,.\vr\\

\hline $2_6$ & $\lan\theta\Dz+\Dt;\,z^i\theta\ran$ &
$\ep=0\,,\;\al=\frac 12\,,\;s=m,m+1$\,.\vr\\

\hline $2_7$ & $\lan\theta z\Dz+\Dt+\de\theta,\,\theta\Dz;\,z^i\theta\ran$ &
$\ep=\al=0$\: and either:\vr\\
& & \qquad {\em i)\:} $s=m$\,,\; $\de=\be$\,.\\
& & \qquad {\em ii)\:} $s=m+1=0$\,.\vr\\
\end{tabular}
\end{center}}\vspace{.5cm}
\footnotesize{In Cases $2_3$--$2_7$ the index $i=0,\ldots,m$.}
\end{table}

\begin{table}[h]
\caption{Odd subspaces for Lie algebras of type $\fse^3$.}
\small{
\begin{center}
\begin{tabular}{c|l|l} Label & \multicolumn{1}{c|}{$\fso$} &
\multicolumn{1}{c}{Rules}\vr\\

\hline\hline $3_1$ & $\lan \theta z^{i-1}(z\Dz+2\be i/m)\ran$ &
$\al=1-\frac m2$, and $\be=0$ if $m=0$\,.\vr\\

\hline $3_2$ & $\lan z^i\theta;\,\theta z^{m+1}(z\Dz+2\be),\,\theta z^j\Dz\ran$
& $2\al=-m$\,;\quad $j=0,\ldots,m+1\,.$\vr\\

\hline $3_3$ & $\lan z^i\Dt\ran$ & $2\al=m$\,.\vr\\

\hline $3_4$ & $\lan\Dt+\ep^\ast\hat\ep\theta;\,
\hat\ep\theta,\,\theta z^{j-1}(z\Dz+j\be)\ran$
& $\al=0$\,;\quad $j=0,1,2$\,.\vr\\
& & $1\in\fse^3$ if $\hat\ep=1$\,.\vr\\

\hline $3_5$ & $\lan\theta z\Dz+\ep^\ast z\Dt+2\be\theta,\, \theta\Dz+\ep^\ast\Dt,\,
\ep z\Dt,\,\ep\Dt\ran$ & $2\al=1$; $\ga=-2\be$ if $\ep=1$\,.\vr\\

\hline $3_6$ & $\lan\theta\Dz,\,z^j\Dt\ran$ & $\al=1$; $\be=0$; $j=0,1,2$\,.\vr\\

\end{tabular}
\end{center}}\vspace{.5cm}
\footnotesize{The index $i$ takes values $i=0,\ldots,m$.}
\end{table}

\section{The QES Lie superalgebras}

In this section we will determine which of the Lie superalgebras $\fs$ obtained in
Section 5 (whose odd subspace $\fso$ is nonzero) are QES, i.e. admit a
nontrivial \fdim{} module $\fn\subset\C(\La^1)$, and will classify all such
modules. We start with the following elementary result:
\begin{lemma}
Let $\fs$ be a \fdim{} graded subalgebra of $\fd$ with a nontrivial odd subspace $\fso$.
Then $\fs$ admits a non-zero \fdim{} module of functions $\fn\subset\C(\La^1)$ if and
only if $\fse\cap\fa\subset\fa_0$.
\end{lemma}
We shall denote the most general $\lan 1,\theta\ran$-translation bimodule
by
$$
\fn(N)=\bigoplus \begin{Sb} \nu\in N \\ i=1,2,3 \end{Sb}
                  \fn^i_\nu\,e^{\nu z}\,,
$$
where $N$ is a finite collection of complex numbers, and
\begin{gather}\label{n123}
\fn^1_\nu=\big\lan z^k\:|\:0\leq k\leq p_\nu\big\ran\,,\qquad
\fn^2_\nu=\big\lan z^k\theta\:|\:0\leq k\leq q_\nu\big\ran\,,\notag\\
\fn^3_\nu=\big\lan z^{p_\nu+k}+z^{q_\nu}\sum_{j=1}^k c^{\nu,k}_j\,z^j\,\theta\,
                  \:|\:1\leq k\leq t_\nu\big\ran\,.
\end{gather}
As in Section 4,
$$
c^{\nu,k}_j=\frac{(p_\nu+k)!(q_\nu+1)!}{(q_\nu+j)!(p_\nu+k-j+1)!}\,c^{\nu,k-j+1}\,,
$$
where $c^{\nu,1},\ldots,c^{\nu,t_\nu}\in\CC$, and $c^{\nu,1}\neq 0$.
We shall also consider the set
$$
\fn_0=\fn(\{0\})\,,\quad\text{with}\quad c^{0,k}=\de_{k1}\,c\,.
$$

In Table 5, we present the list of the even QES subalgebras
$\fse^i$ in Corollary~\ref{cor.even}. The calculations needed to complete
Table~5 present no difficulties.
In Tables~6--9, we present the possible odd subspaces $\fso$ corresponding to
the even QES subalgebras in Table~5, along with the invariant modules of
functions for
$\fs=\fse\oplus\fso$. The image of a module $\fn$ under the action of the
elements of $\fso$ will be denoted by $\fso[\fn]$.

The case $1_4$ in Table~7
perhaps deserves some special attention.
According to Table~5, we have $\Si=\emptyset,\{1\}$. It follows from
Table~2 that $\fso=\lan \theta\Dz+\Dt; \theta\ran$, and $\theta\in\fso\Ra 1\in\fse^1$.
Assume first that $\theta\notin\fso$. It is easy to see that $p_\nu=q_\nu$
in equation~\eqref{n123} for $\nu\neq 0$. Remarkably, for $\nu\neq 0$ the parameters
$c^{\nu,l}$, $l=1,\ldots,t_\nu$, defining the {\em mixed} functions in $\fn^3_\nu$
are no longer arbitrary but of the form
$$
c^{\nu,l}=\frac{(q_\nu+l)!}{(q_\nu+1)!}\nu^{\frac 32-l}a_l\,,
$$
where $a_l$ are fixed constants that can be determined recursively from the
condition
$(\theta\Dz+\Dt)[\fn^3_\nu]\subset\fn^3_\nu/\fn^2_\nu$. The first ten
constants are
$$
a_l=1,\frac 12,\frac 1{2^3},\frac 1{2^4},\frac 5{2^7},\frac 7{2^8},
\frac{3\cdot 7}{2^{10}},\frac{3\cdot 11}{2^{11}},\frac{3\cdot 11\cdot 13}{2^{15}},
\frac{5\cdot 11\cdot 13}{2^{16}},\ldots
$$
Finally, it is easy to see that $p_0=q_0,q_0+1$ and $t_0=0$. If $\theta\in\fso$,
then $1\in\fse^1$ and the associated module is $\fn^{1.1}$, with the
constraint $p_\nu=q_\nu$ for all $\nu\in N$.

The parameters $\al$ and $\be$ appearing
in the superalgebras of type $\fs^3$ are quantized. This phenomenon,
the {\em quantization of the cohomology}, has been noticed before in the
case of \fdim{} QES Lie algebras and has received some study,
\cite{GHKO93}, \cite{PV96}. The parameter $\al$ is quantized for all \fdim{}
Lie superalgebras of the type $\fs^3$, whereas $\be$ is only quantized 
in every case after imposing the QES condition. Finally, it is also worth mentioning that 
the Lie superalgebra $\osp{2,2}\iso\fsl(2|1)$, extensively used in the
construction of QES matrix Hamiltonians (see for instance \cite{ST89},
\cite{BK94}, \cite{FGR96}), is the QES superalgebra $3_5$ in Tables 4 and 9
with $\ep=1$.

\begin{table}[h]
\caption{Even QES subalgebras $\fse$ and associated modules $\fn$.}
\small{
\begin{center}
\begin{tabular}{l|l|l} \multicolumn{1}{c|}{$\fse$} & \multicolumn{1}{c|}{$\fn$} &
\multicolumn{1}{c}{Rules}\vr\\

\hline\hline $\fse^0=\lan\ep\theta\Dt+\al;\,1\ran$
& $\fn^{0.0}=\lan f_j+g_j\theta\ran$ & $\ep=\al=0$\,.\vr\\
\cline{2-3} & $\fn^{0.1}=\lan f_j,\,g_k\theta\ran$ & $\ep=1$\,.\vr\\

\hline $\fse^1=\lan\Dz,\,\ep\theta\Dt+\al;\,1\ran$ & $\fn^{1.0}=\fn(N)$ & $\ep=\al=0$\,.\vr\\
\cline{2-3} & $\fn^{1.1}=\bigoplus_{\nu\in N}\fn^1_\nu\oplus\fn^2_\nu$ & $\ep=1$\,.\vr\\

\hline $\fse^2=\lan\Dz,\,z\Dz+\al\theta\Dt+\be;\,\ep\theta\Dt+\ga,\,1\ran$
& $\fn^{2.0}=\fn_0$ & $\ep=0$\,;\;$t\geq 1\Ra\al=p-q$\vr\,.\\
\cline{2-3} & $\fn^{2.1}=\lan z^j,\,z^k\theta\ran$ & $\ep=1$\,.\vr\\

\hline $\fse^3=\lan\Dz,\,z\Dz+\al\theta\Dt+\be,\,z^2\Dz$
& $\fn^{3.0}=\lan z^j(1+c\theta)\ran$ & $\ep=\al=0$\,;\;$2\be=-p$\,.\vr\\
\cline{2-3} $\qquad\qquad +2\al z\theta\Dt+2\be z;\,\ep\theta\Dt+\ga,\,1\ran$
& $\fn^{3.1}=\lan z^j\ran$ & $2\be=-p$\,.\vr\\
\cline{2-3} & $\fn^{3.2}=\lan z^k\theta\ran$ & $2(\al+\be)=-q$\,.\vr\\
\cline{2-3} & $\fn^{3.3}=\lan z^j,\,z^k\theta\ran$ & $2\be=-p$\,;\;$2(\al+\be)=-q$\,.\vr\\

\end{tabular}
\end{center}}\vspace{.5cm}
\footnotesize{The indices $j,k$ take values $j=0,\ldots,p$, $k=0,\ldots,q$}
\end{table}

\begin{table}[h]
\caption{Odd subspaces $\fso$ for even QES Lie algebras $\fse^0$ and $\fs$-modules $\fn$.}
\small{
\begin{center}
\begin{tabular}{c|l|l|l} \multicolumn{1}{c|}{Label} & \multicolumn{1}{c|}{$\fso$}
& \multicolumn{1}{c|}{$\fn$} & \multicolumn{1}{c}{Rules}\vr\\

\hline\hline $0_1$ & $\lan\theta(\phi_i\Dz+\om_i)\ran$
& $\fn^{0.\ep}+\fso[\fn^{0.\ep}]$ & \vr\\

\hline $0_{2a}$ & $\lan\chi_i\Dt\ran$ & $\fn^{0.\ep}+\fso[\fn^{0.\ep}]$ & \vr\\

\hline $0_{2b}$ & $\lan\Dt+\hat\ep^\ast\theta;\,\hat\ep\theta\ran$
& $\lan f_j+g_j\theta,\,g_j+f_j\theta\ran$
& $\hat\ep=\ep=0$\,,\quad $1\in\fse^0$\,.\vr\\
\cline{3-4} & & $\lan f_j,f_j\theta\ran$ & $\hat\ep=1$\,,\quad $1\in\fse^0$\,.\vr\\

\end{tabular}
\end{center}}\vspace{.5cm}
\footnotesize{The indices $i,j$ take values $i=1,\ldots,m$, $j=1,\ldots,p$}
\end{table}

\begin{table}[h]
\caption{Odd subspaces $\fso$ for even QES Lie algebras $\fse^1$ and $\fs$-modules $\fn$.}
\small{
\begin{center}
\begin{tabular}{c|l|l|l} \multicolumn{1}{c|}{Label} & \multicolumn{1}{c|}{$\fso$}
& \multicolumn{1}{c|}{$\fn$} & \multicolumn{1}{c}{Rules}\vr\\

\hline\hline $1_1$ & $\fm(\lan\theta\Dz,\theta\ran,M)$
& $\fn^{1.\ep}+\fso[\fn^{1.\ep}]$ & \vr\\

\hline $1_{2a}$ & $\lan z^i e^{\mu z}\Dt\ran$ & $\fn^{1.\ep}+\fso[\fn^{1.\ep}]$
& $0\leq i\leq m_\mu\,,\quad\mu\in M$\,.\vr\\

\hline $1_{2b}$ & $\lan\Dt+\hat\ep\theta;\,\hat\ep^\ast\theta\ran$
& $\fn^{1.\hat\ep^\ast}$ & $1\in\fse^1$, and $\ep=1\Ra\hat\ep=0$\,.\vr\\
& & & $p_\nu=q_\nu$\,;\quad $c^{\nu,k}=\de_{k,1}$\,.\vr\\

\hline $1_3$ & $\lan\theta\Dz,\,\Dt+\hat\ep\theta;\,\theta\ran$
& $\lan\fn^{1.1};\,\tilde\ep z^{q_0+1}(1+\hat\ep\theta)\ran$
& $\theta\in\fso\Ra 1\in\fse^1\;\text{and}\;\tilde\ep=0$\,.\vr\\
& & & $\hat\ep=1\Ra 1\in\fse^1\;\text{and}\;\ep=0$\,.\\
& & & $p_\nu=q_\nu$\,.\vr\\

\hline $1_4$ & $\lan\theta\Dz+\Dt;\,\theta\ran$
& $\fn^{1.0}$ & $\theta\notin\fso$. If $\nu\neq 0$, we have:\vr\\
& & & \qquad {\em i)\:} $p_\nu=q_\nu$\,.\\
& & & \qquad {\em ii)\:} $c^{\nu,l}=\frac{(q_\nu+l)!}{(q_\nu+1)!}\nu^{\frac 32-l}a_l$\,, with\vr\\
& & & \qquad\quad $a_l$ fixed constants and $l=1,\ldots,t_\nu$\,.\\
& & & $p_0=q_0,\,q_0+1$;\quad $t_0=0$\,.\vr\\
\cline{3-4} & & $\fn^{1.1}$ & $\theta\in\fso^1$\,;\quad $1\in\fse^1$\,;\quad $p_\nu=q_\nu$\,.\vr\\

\end{tabular}
\end{center}}\vspace{.5cm}
\footnotesize{The parameter $\ep=0$ in Case $1_4$.}
\end{table}

\begin{table}[h]
\caption{Odd subspaces $\fso$ for even QES Lie algebras $\fse^2$ and $\fs$-modules $\fn$.}
\small{
\begin{center}
\begin{tabular}{c|l|l|l} \multicolumn{1}{c|}{Label} & \multicolumn{1}{c|}{$\fso$}
& \multicolumn{1}{c|}{$\fn$} & \multicolumn{1}{c}{Rules}\vr\\

\hline\hline $2_1$ & $\fm_0(\lan\theta\Dz,\theta\ran)$
& $\fn^{2.\ep}+\fso[\fn^{2.\ep}]$ & $r\geq 1\Ra m=n+1$\vr\\

\hline $2_{2a}$ & $\lan z^i\Dt\ran$ & $\fn^{2.\ep}+\fso[\fn^{2.\ep}]$
& $0\leq i\leq m$\,.\vr\\

\hline $2_{2b}$ & $\lan\Dt+\hat\ep\theta;\,\hat\ep^\ast\theta\ran$
& $\fn^{2.\hat\ep^\ast}$ & $1\in\fse^2$, and $\hat\ep=1\Ra\ep=\al=0$\,.\vr\\
& & & $p=q$\,;\quad $c=1$\,.\vr\\

\hline $2_3$ & $\lan\theta\Dz,\,\Dt+\hat\ep\theta;\,\theta\ran$
& $\lan\fn^{2.1};\,\tilde\ep z^{q+1}(1+\hat\ep\theta)\ran$
& $\theta\in\fso\Ra 1\in\fse^2\;\text{and}\;\tilde\ep=0$\,.\vr\\
& & & $\hat\ep=1\Ra 1\in\fse^2\;\text{and}\;\ep=\al=0$\,.\\
& & & $p=q$\,.\vr\\

\hline $2_4$ & $\lan\theta(z\Dz+\de),\,\theta\Dz,\,\Dt+\ep^\ast\hat\ep\,\theta;\,
\theta\ran$ & $\lan\fn^{2.1};\,\tilde\ep z^{q+1}(1+\ep^\ast\hat\ep\theta)\ran$
& $\al=0$\,,\quad $p=q$\,.\vr\\
& & & $1\notin\fse^2\Ra\theta\notin\fso$, $\de=\be$, $\hat\ep=0$\,.\\
& & & $\tilde\ep=1\Ra\theta\notin\fso$, $\de=-(q+1)$\,.\vr\\

\hline $2_5$ & $\lan\theta\Dz+\hat\ep^\ast z\Dt,\,\hat\ep z\Dt,\,\Dt\ran$
& $\fn^{2.1}$ &  $p=q+1$ and either:\vr\\
& & & \qquad {\em i)\:} $\ep=\be=0$\,,\; $\al=1$\,.\\
& & & \qquad {\em ii)\:} $\ep=\hat\ep=1$\,,\; $\be=-\ga$\,.\vr\\

\hline $2_6$ & $\lan\theta\Dz+\Dt;\,\theta\ran$ & $\fn^{2.1}$
& $\theta\in\fso\Ra 1\in\fse^2$ and $p=q$\,.\vr\\
& & & $\theta\notin\fso\Ra p=q,q+1$\,.\vr\\

\hline $2_7$ & $\lan\theta z\Dz+\Dt+\de\theta,\,\theta\Dz;\,\theta\ran$
& $\lan\fn^{2.1};\,\tilde\ep z^{q+1}(1$ & $\ep=\al=0$\,,\quad $p=q$\,.\vr\\
& & $\qquad+(\de+q+1)^{\frac 12}\theta)\ran$
& $1\notin\fse^2\Ra\theta\notin\fso$ and $\de=\be$\,.\\
& & & $\theta\in\fso\Ra\tilde\ep=0$\,.\vr\\

\end{tabular}
\end{center}}\end{table}

\begin{table}[h]
\caption{Odd subspaces $\fso$ for even QES Lie algebras $\fse^3$ and $\fs$-modules $\fn$.}
\small{
\begin{center}
\begin{tabular}{c|l|l|l} \multicolumn{1}{c|}{Label} & \multicolumn{1}{c|}{$\fso$}
& \multicolumn{1}{c|}{$\fn$} & \multicolumn{1}{c}{Rules}\vr\\

\hline\hline $3_1$ & $\lan\theta z^{i-1}(z\Dz+2\be i/m)\ran$
& $\lan 1+\theta\ran$ & $\ep=\be=0$, $m=2$\,.\vr\\
\cline{3-4} & & $\lan 1\ran$ & $\be=0$\,.\vr\\
\cline{3-4} & & $\fn^{3.2}$ & $m\geq 1$; $2\be=m-q-2$\,.\vr\\
\cline{3-4} & & $\fn^{3.3}$ & $m=1,2$; $2\be=m-q-2=-p$\,.\vr\\

\hline $3_2$ & $\lan z^i\theta;\,\theta z^{m+1}(z\Dz+2\be),\,\theta z^j\Dz\ran$
& $\fn^{3.2}$, $\fn^{3.3}$ & $2\al=-m$\,;\quad $j=0,\ldots,m+1\,.$\vr\\
& & & $2\be=m-q=-p$\,.\vr\\

\hline $3_3$ & $\lan z^i\Dt\ran$ & $\fn^{3.1}$, $\fn^{3.3}$
& $2\al=m$; $-2\be=p=m+q$\,.\vr\\

\hline $3_{4a}$ & $\lan\Dt;\,\theta,\,\theta z^{j-1}(z\Dz+j\be)\ran$ & $\fn^{3.1}$
& $\theta\notin\fso$; $2\be=-p$\,.\vr\\
\cline{3-4} & & $\fn^{3.3}$ & $2\be=-p=-q$, and $\theta\in\fso\Ra 1\in\fse^3$\,.\vr\\

\hline $3_{4b}$ & $\lan\Dt+\theta;\,\hat\ep\theta z^{j-1}(z\Dz+j\be)\ran$
& $\fn^{3.0}$ & $c=1$; $2\be=-\hat\ep^\ast p$\,.\vr\\
\cline{3-4} & & $\fn^{3.3}$ & $2\be=-p=-q$\,.\vr\\

\hline $3_5$ & $\lan\theta z\Dz+\ep^\ast z\Dt+2\be\theta,$ & $\lan 1\ran$
& $2\al=1$; $\be=0$\,.\vr\\
\cline{3-4} & $\qquad\theta\Dz+\ep^\ast\Dt,\,\ep z\Dt,\,\ep\Dt\ran$
& $\fn^{3.3}$ & $2\al=1$; $-2\be=p=q+1$\,.\vr\\

\hline $3_6$ & $\lan\theta\Dz,\,z^j\Dt\ran$ & $\lan 1\ran$
& $\al=1$; $\be=0$; $j=0,1,2$\,.\vr\\

\end{tabular}
\end{center}}\vspace{.5cm}
\footnotesize{The index $i$ takes values $i=0,\ldots,m$.
In Case $3_1$, $\al=1-\frac m2$ and $\be=0$ if $m=0$.
In Cases $3_4$, $\al=0$ and $j=0,1,2$. In Case $3_{4b}$, we also have $\ep=0$
and $1\in\fse^3$. In Case $3_5$, $\ga=-2\be$ if $\ep=1$.}
\end{table}

% theend

\end{document}